\documentclass[12pt]{article}
\usepackage[T1]{fontenc}
\usepackage{graphicx}
\usepackage{amsfonts}
\usepackage{xcolor}
\usepackage{comment}
\usepackage{amssymb}
\usepackage[bookmarksopen=true]{hyperref}
\usepackage{bookmark}
\newtheorem{lemma}{Lemma}
\newtheorem{theorem}{Theorem}
\newtheorem{corollary}{Corollary}
\usepackage{microtype} 
\usepackage{mismath, algpseudocode, amsfonts, hyperref, algorithm}

\begin{document}
	\title{Maximizing Weighted Dominance in the Plane}
	
	\author{Waseem Akram and
		Sanjeev Saxena\\
	Dept. of Computer Science and Engineering, \\
		Indian Institute of Technology, Kanpur, INDIA-208 016\\
		Email: {\{akram,ssax\}@iitk.ac.in}}
	\maketitle              
	\begin{abstract}
		Let $P$ be a set of $n$ weighted points, $Q$ be a set of $m$ unweighted points in the plane, and $k$ a non-negative integer. We consider the problem of computing a subset $Q'\subseteq Q$ with size at most $k$ such that the sum of the weights of the points of $P$ dominated by at least one point in the set $Q'$ is maximized. A point $q$ in the plane dominates another point $p$ if and only if $x(q)\ge x(p)$ and $y(q)\ge y(p)$, and at least one inequality is strict.
		
		We present a solution to the problem that takes $O(n + m)$-space and 
		$O(k\min\{n+m, \frac{n}{k}+m^2\}\log m)$-time. We (conditionally) improve upon the existing result (the bounds of our solution are interesting when $m= o(\sqrt{n}))$.
		Moreover, we also present a simple algorithm solving the problem in $O(km^2+n\log m)$-time and $O(n+m)$-space. 
		The bounds of the algorithm are interesting when $m= o(\sqrt{n})$.
		
	\end{abstract}
	\section{Introduction}
	
	In the $d$-dimensional Euclidean space, a point $q$ dominates another point $p$ if and only if the coordinate of $q$ is greater than or equal to that of $p$ in all dimensions and strictly greater in at least one dimension \cite{Tao09}. In this paper, we study the following problem.
	\begin{quote}
		Given a set $P$ of $n$ (possibly negative) weighted points, a set $Q$ of $m$ unweighted points in the plane, and a non-negative integer $k$, find a subset $Q'\subseteq Q$ of size at most $k$ such that the sum of weights of the points of $P$ that are dominated by some point(s) in $Q'$ is maximized.
	\end{quote}
	Choi, Cabello, and Ahn \cite{Choi21} introduced the problem and called it \textit{maxDominance} problem. 
	They also gave a solution which uses $(m+n)$-space and $O(k(m+n)\log m)$-time.
	The problem has applications in multi-criteria decision-making scenarios \cite{Cabello23,Lin06,Li19,Tao09,Emmerich18,Mukh14}.
	
	We conditionally improve the bounds of the solution given in \cite{Choi21} by employing a preprocessing step. We show that the \textit{maxDominance} problem can be solved using $O(m + n)$-space and $O(k\min\{n+m, \frac{n}{k}+m^2\}\log m)$-time. The bounds are interesting when $m= o(\sqrt{n})$.
	For the case when $m=\Omega(\sqrt{n})$, the bounds become the same as that of Choi et al.'s \cite{Choi21}.
	
	We also present a simple algorithm to solve the maxDominance problem in $O(km^2+n\log m)$-time and $O(n+m)$-space. The time bound is interesting only when $m= o(\sqrt{n})$. The algorithm uses no advanced data structures (unlike the use of segment tree in \cite{Choi21}).
	
	\paragraph{Motivation}:
	A point in a set of points in $\mathbb{R}^d$ is \textit{maximal} if no other point dominates it. A maximal point is also called \textit{skyline point}. Computation of all skyline points is an important and well-studied problem in computational geometry \cite{Cabello23,Chan96,Cormen09}. The problem finds applications in numerous domains, including database management and multi-criteria decision-making \cite{Cabello23,Choi21,Chan96,Godfrey05,Li19,Mukh14}. 
	However, in the presence of many skyline points, reporting the whole skyline may not be very helpful; a user may find it challenging to understand the possible trade-offs the complete skyline offers.
	A possible resolution to this issue is to find and report a subset of skyline points that best ``represent'' them instead of computing the entire skyline. Such a subset is called \textit{representative skyline} \cite{Tao09}.
	
	\paragraph{Previous and Related Works}:
	The problem of computing representative skyline has been studied in different settings \cite{Cabello23,Choi21,Lin06,Tao09}. Tao et al. \cite{Tao09} defined a distance-based representative skyline that minimizes the Euclidean distance between a non-representative point and its nearest representative skyline point. In $2$-dimensional space, they gave an algorithm to compute such a representative skyline in $O(n\log h + kh^2)$ time, where $n$ is the size of the input set and $h$ is the number of skyline points. 
	Recently, Cabello \cite{Cabello23} reduced the running time to $O(n\log h)$.
	
	Lin et al. \cite{Lin06} studied the problem of computing $k$ skyline points so that the number of points dominated by at least one of the $k$ points is maximized. They present a dynamic algorithm that solves the problem in $O(kh^2+n\log h)$ time in the plane. Here, $n$ and $h$ are the numbers of input and skyline points, respectively. Choi, Cabello, and Ahn \cite{Choi21} considered a more general problem where a (possibly negative) weight is assigned with each point, namely \textit{the maxDominance} problem (defined earlier in the section). They solved the problem using $O(k(n+m)\log m)$ time and $O(n+m)$ space. For the case when the points in $P$ are allowed to have positive weights only, they simplified the algorithm, which uses $O(m\log s + k(n+s)\log s)$ time and $O(n+m)$ space, where $s$ is the number of skyline points in $Q$.
	Note that the problem studied by Lin et al. is a particular case of the maxDominance problem: assigning unit weight to every point in $P$ and setting $Q$ to be the skyline points.

	\subsection{Definitions and Notations}
	We use the notations from \cite{Choi21}. Let $P$ be a set of $n$ weighted points, $Q$ be a set of $m$ unweighted points in the plane, and $k$ a positive integer. The weight of a point $p$, denoted by $w(p)$, could be negative.
	We denote by $x(p)$ and $y(p)$ the $x$-coordinate and $y$-coordinate of a point $p$ in the plane, respectively. Let $x_{min}$ and $x_{max}$ be the minimum and the maximum $x$-coordinates in the set $P\cup Q$, respectively. Similarly, $y_{min}$ and $y_{max}$ are defined.
	Let $Q=\{q_1, q_2,..., q_m\}$ be sorted in decreasing order of $y$-coordinates. 
	The set of the points $p\in P$ (strictly) lying above the horizontal line through 
	point $q_i$ is denoted by $P_i$. Formally, $P_i=\{p\in P: y(p)>y(q_i)\}$
	
	For a given region $R\in \mathbb{R}^2$, the weight of region $R\in \mathbb{R}^2$ 
	is defined as the sum of the weights of all points of $P$ contained in the region 
	$R$. It is denoted by $w(R)$. If no point from the set $P$ lies in $R$, then $w(R)$ 
	is defined as zero. For any point $q$ in the plane, $dom(q)$ represents the region 
	$(-\infty, x(q)] \times (-\infty, y(q)]$. For a set $Q$ of points, 
	$dom(Q)$ is defined as the union of the regions $dom(q), q\in Q$. Let $[t]$ denotes the set of natural numbers $\{1,2,....,t\}$.
	
	In the maxDominance problem, we are to find a subset $Q'\subseteq Q$ with size at most $k$ such that $w(dom(Q'))$ is maximized. Formally, we want to compute
	maxDominance$(P,Q,k)=\max \{w(dom(Q')): Q'\subseteq Q, |Q'|\le k\}$ and to obtain a subset $Q'\subseteq Q$ with size at most $k$ and $w(Q') = $ maxDominance$(P,Q,k)$.
	
	In Section~$2$, we describe an algorithm that conditionally improves the bounds of Choi et al.'s \cite{Choi21}. A simple algorithm to solve the maxDominance problem is presented in Section~$3$. We conclude our work in Section~$4$.
	
	\section{Transformation of Input points}
	In this section, we conditionally improve the bounds of the algorithm given by Choi, Cabello, and Ahn \cite{Choi21}.
	We define a set $P'$ of weighted points, whose size is bounded above by $\min\{n, m^2\}$, such that maxDominance$(P,Q,k)$ = maxDominance$(P',Q,k)$. We show that the set $P'$ can be computed in $O((m+n)\log m)$-time and $O(m+n)$-space. We then run the Choi et al.'s algorithm \cite{Choi21} on the new instance.

	The points of the set $P$ present outside the region $dom(Q)$ do not contribute to any solution (and can be easily removed in a preprocessing step). Thus, we need to consider only the points of $P$ present in the region $dom(Q)$.
	We partition the region $dom(Q)$ into several disjoint rectangles (also called \textit{cells}).
	Consider the horizontal lines through each point $q\in Q$ and the vertical (downward) rays from the points of the set $Q$. 
	The intersections of these objects are the vertices of the cells. Each cell containing a point $p\in P$ would correspond to a point of $P'$. An important property of the partition is that the points in the same cell are dominated by the same set of points of $Q$.
	The following paragraph describes an algorithm computing the set $P'$ efficiently.

	We now describe how to compute the set $P'$ efficiently.
	We assume that no two points of $Q$ have the same $x$- or $y$-coordinate:
	the points in the set $Q$ can be transformed so that no two (transformed) points have the same $x$- or $y$-coordinate and $dom(q)\cap P = dom(q')\cap P$, where $q'$ is the transformed point of $q\in Q$. This step can be done in $O((n+m)\log m)$ time \cite{Choi21}.
	Let the set $Q=\{q_1, q_2,...., q_m\}$ be sorted in decreasing order of the $y$-coordinates.
	Consider a partition of the plane induced by horizontal and vertical lines through the points of $Q$. The plane will be divided into a collection of $(m+1)^2$ non-overlapping rectangles, which we will call cells.
	The cells between the horizontal lines through $q_{i}$ and $q_{i+1}$, for any $i\in [m]$, are characterized as follows: Let $\pi$ be a permutation of $[i]$ such that $q_{\pi(1)}, q_{\pi(2)}, ..., q_{\pi(i)}$ is sorted in increasing order of $x$-coordinates.
	For every $j\le i$, we denote the cell with lower-left corner $(x(q_{\pi(j-1)}), y(q_{i+1}))$ and upper-right corner $(x(q_{\pi(j)}), y(q_{i}))$ by $C_{ij}$. Here, $q_0=(x_{min}-1,y_{max}+1)$ and $q_{m+1}=(x_{max}+1, y_{min}-1)$ with $q_0=q_{\pi(0)}$ and $q_{m+1}=q_{\pi_(m+1)}$.
	See Figure~$1.$
	\begin{figure}[h]
		\centering
		\includegraphics[scale=.16]{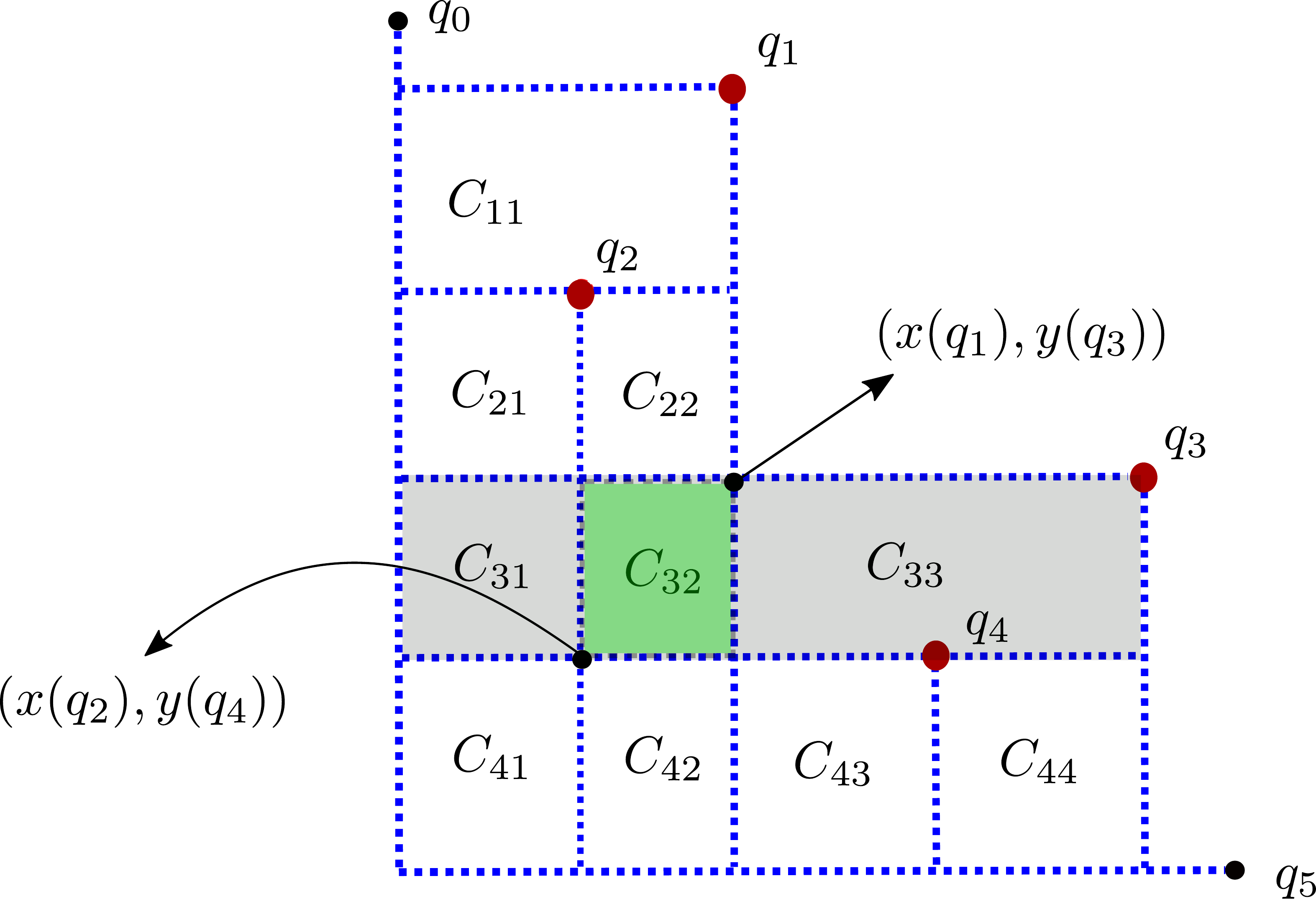}
		\caption{ Cells between lines through $q_3$ and $q_4$ are shaded with grey. The cell $C_{32}$ has been shown in green shade along with corner points.}
	\end{figure}
	We call a cell empty if it contains no point from the set $P$.
	The points in non-empty cells form a partition of set $P$, i.e., no two cells have a point in common, and the union of points of all the non-empty cells gives the set $P$. 
	
Remark: 
		As the total number of cells is $(m+1)^2$. Thus, the number of non-empty cells in the partition is $O(m^2)$.

	\begin{lemma}\label{lem:dom}
		Each pair of points in $P$ contained in a given cell $C_{ij}$ have the same dominating points from the set $Q$.
	\end{lemma}
	Proof:
		For the sake of contradiction, let us assume that two points $p_r$ and $p_s$ contained in the cell $C_{ij}$ do not have the same set of dominating points from the set $Q$. Without loss of generality, let us assume that $y(p_r) > y(p_s)$. Since points $p_r$ and $p_s$ are in the cell $C_{ij}$, so their $x$- coordinates would lie in the range $[x(q_{\pi(j-1)}), x(q_{\pi(j)})]$ and $y$-coordinates in the range $[y(q_i), y(q_{i+1})]$.
		
		Since the points $p_r$ and $p_s$ do not have the same set of dominating points, at least one point, say $q$, of $Q$, would dominate one point but not the other. Let point $q$ dominates $p_r$, but not $p_s$. The other scenario, where $q$ dominates $p_s$, but not $p_r$, can be dealt with analogously. Note that $x(q)$ must be less than $x(p_s)$, otherwise, $q$ would dominate $p_s$ as well. So $x(q)$ is in the range $[x(p_r), x(p_s)]\subseteq [x(q_{\pi(j-1)}), x(q_{\pi(j)})]$. However, by the construction of cells, there can not be a point of $Q$ having $x$-coordinate (or $y$- coordinate) in any cell's $x$-range (resp. $y$-range). This contradiction shows that 
		all the points of set $P$ in $C_{ij}$ have the set of dominating points from the set $Q$.$\square$
	
	We create a weighted point $p$ for each non-empty cell $C_{ij}$ with $w(C_{ij})\ne 0$ such that $p$ lies in $C_{ij}$ and its weight is $w(C_{ij})$. We call the point $p$ the representative of the points in the cell $C_{ij}$. The set $P'$ consists of the representatives of the non-empty cells $C_{ij}$ with $w(C_{ij})\ne 0$. See Figure~2.
	\begin{figure}[h] \label{fig:rep} 
		\centering 
		\includegraphics[scale=.16]{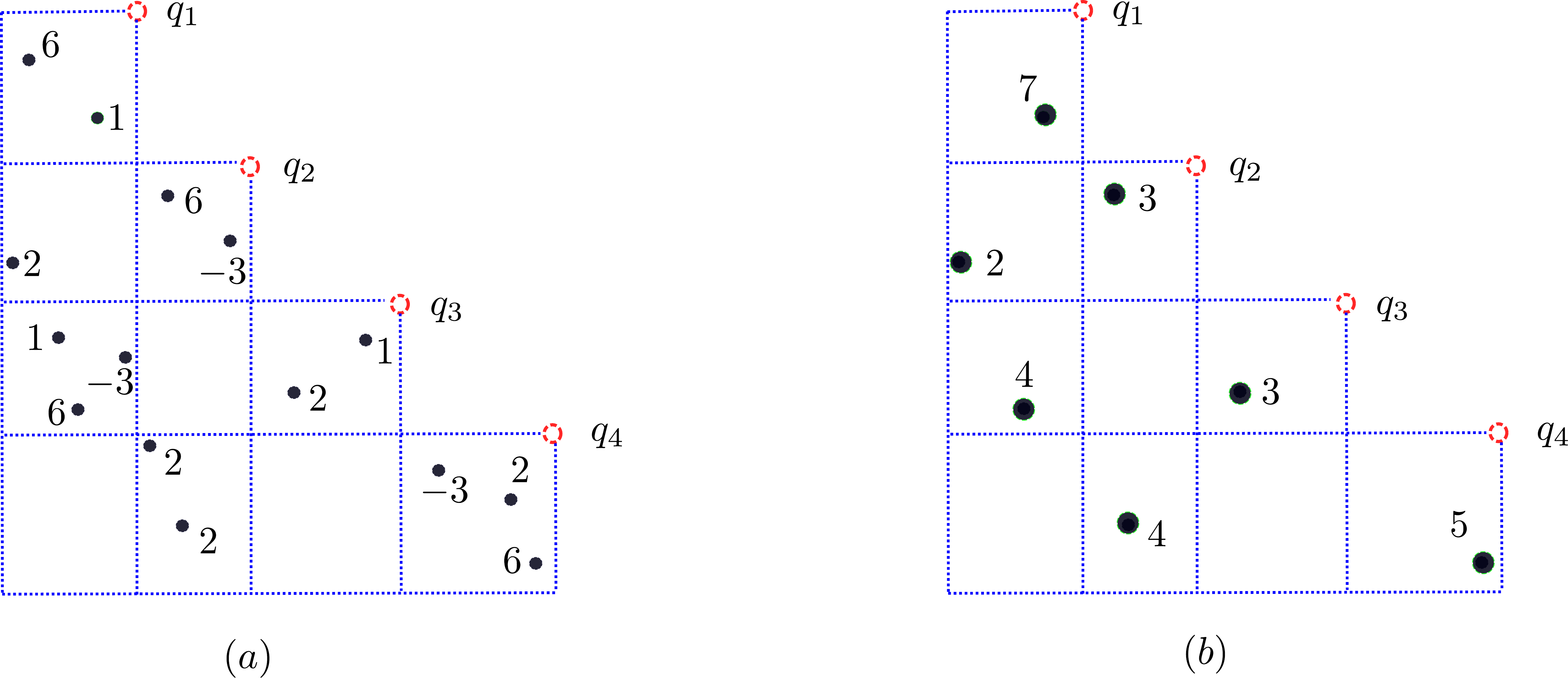}
		\caption{(a) represents the non-empty cells of the partition, and (b) the cells containing their representatives.}
	\end{figure}
	Note that the number of points in $P'$ is at most $\min\{n,m^2\}$. From Lemma~\ref{lem:dom} and the construction of $P'$, we get
	\begin{corollary}\label{cor:equi}
		maxDominance$(P,Q,k)$ = maxDominance$(P',Q,k)$.
	\end{corollary}
	We now describe an algorithm that efficiently computes the representative of 
	each non-empty cell. 
	We store the set $Q$ into two sorted lists $L_x$ and $L_y$. The list $L_x$ is sorted in increasing order of $x$-coordinates, and $L_y$ is sorted in decreasing order of $y$-coordinates. 
	We create a tuple $(i,j)$ for each point $p\in P$, where $i$ is the index of the next smaller $x$-coordinate in $L_x$ and $j$ is the index of the next larger $y$-coordinate in $L_{y}$.
	Let $S$ be the set of all tuples, one for each point in $P$. Thus, $|S|=|P|=n$. Note that a tuple in the set may correspond to more than one point of $P$. All points having the same tuple $(i,j)$ belong to the cell $C_{ij}$. We now sort the set $S$ using the radix sort. In the sorted set, all copies of a tuple would be contiguous. Thus, we can compute the representative of each non-empty cell $C_{ij}$ using a single scan of the sorted set $S$. The algorithm uses $O((n+m)\log m)$-time and $O(n+m)$-space.
	Thus, we have the following lemma. 
	\begin{lemma}\label{lem:rep}
		We can compute the representative of every non-empty cell in $((n+m)\log m)$-time and $O(n+m)$-space.
	\end{lemma}
	Choi et al. \cite{Choi21} gave an efficient dynamic programming-based algorithm for the maxDominance problem. Their main result is as follows.
	
	\begin{theorem}
		(Theorem~$4$ in \cite{Choi21})
		The maxDominance problem in the plane for $n$ weighted points in $P$
		and $m$ points in $Q$ using at most $k$ points of $Q$ can be solved in $O(k(n + m) \log m)$ time using $O(n + m)$ space. The result also holds when there are negative weights.
	\end{theorem}
	We now employ Choi et al.'s algorithm \cite{Choi21} to solve the new instance $(P', Q,k)$ of the maxDominance problem. As $|P'|=\min\{n,m^2\}$, we have the following result.
	\begin{theorem}
		Given a set of $P$ of $n$ weighted points, a set $Q$ of $m$ unweighted points, and a positive integer $k$, then we can solve the maxDominance problem in $O(k\min\{n+m, \frac{n}{k}+m^2\}\log m)$ time and $O(n + m)$ space. $\square$
	\end{theorem}
	Remark: 
		In the worst case, the bounds of the solution reduces to that of Choi et al. \cite{Choi21}. The bounds are interesting when $m= o(\sqrt{n})$.

	\section{An $O(km^2+n\log m)$-Time Solution to the maxDominance Problem}
	In this section, we present an algorithm for solving the maxDominance problem in $O(km^2+n\log m)$ time and $O(n+m)$ space. Our approach is similar to that of Choi et al. \cite{Choi21}. In the $l^{th} (1\le l \le k)$ iteration, we compute the value of an optimal solution to the problem with size at most $l$ using (some) values computed in the $(l-1)^{th}$ iteration. Thus, at the end of the $k^{th}$ iteration, we will have the value of an optimal solution to our main problem.
	
	We add a point $q_{m+1}=(x_{max}+1, y_{min}-1)$ to the set $Q$. Note that no point in $P\cup Q$ dominates the point $q_{m+1}$.
	We denote by $T_l(i)$ the value of an optimal solution with size at most $l$ to the maxDominance problem when restricted to $P_i$, and $Q=\{q_1, q_2,..., q_i\}$. 
	Recall that $P_i$ is the set of points of $P$ with $y$-coordinates greater than $y(q_i)$.
	Note that $T_k(m+1)$ corresponds to the value of an optimal solution to the maxDominance$(P, Q, k)$ problem, and $T_{0}(i)=0$ for all $i\in [m+1]$. For all $i,j\in[m+1]$ with $j<i$, $\rho(i,j)$ is defined to be the sum of the weights of the points in $P_i$ that are dominated by point $q_j$. If no such point exists, $\rho(i,j)$ is taken as $0$. For every $i\in[m]$, $\rho(i,i)$ is assumed to be $0$.
	We can compute $\rho(i,j)$ for all $j<i$ as follows: after computing the representatives of the cells $C_{ij}$, $i\in[m]$ and $j\le i$, as described in the previous section, we use the following relation to compute $\rho(i,j)$ values.
	\begin{equation}\label{equ:rho}
	\rho(i,j)=\rho(i-1,j)+\sum_{t=1}^{\pi(j)}w(C_{(i-1)t})
	\end{equation}
	For each $i\in[m]$, we compute prefix sums $psum(i,l)=\sum_{t=1}^{l}w(C_{it})$, for all $l\le i$; the total time used for computing all the prefix sums is $O(m^2)$. 
	Using the prefix sums and the recurrence relation~$(1)$, we then compute $\rho(i,j)$ for all $j<i$. From Lemma~\ref{lem:rep}, $O(m^2 + n\log m)$ time is used for computing all $\rho$ values.
	The space used is $O(n+m^2)$; the factor $O(m^2)$ is for storing all $\rho$ values.
	
	Let us assume that we have all values $T_{l-1}(i)$ for all $i\in [m+1]$. We compute all $T_l(i)$ as follows: 
	We process points $q_1, q_2,...q_{m+1}$ one by one. Let $q_i$ be the current point being processed. We find a point $q_j$ among the points $q_1, q_2,...,q_{j-1}$
	with $x(q_j)\le x(q_i)$ such that the sum $T_{l-1}(j)+\rho(i,j)$ is maximized. A pseudo-code of the algorithm is described in Algorithm~\ref{alg:simple}.
	\begin{algorithm}
		\caption{An $O(km^2+n\log m)$-time algorithm solving \textit{maxDominance} problem}\label{alg:simple}
		\begin{algorithmic}[1]
			\State $x_{max} \gets $ the maximum $x$-coordinate among the points of $P\cup Q$.
			\State $y_{min} \gets $ the minimum $y$-coordinate among the points of $P\cup Q$.
			\State $q_{m+1} \gets (1+x_{max}, -1+y_{min})$
			\State $Q'\gets Q\cup \{q_{m+1}\}$
			\State Let the set $Q'=\{ q_1, q_2, q_3,...., q_{m+1}\}$ be sorted in decreasing order of the $y$-coordinates.
			\State compute $\rho(i,j)$ for every $i,j\in[m+1]$ with $j<i$
			\State $T_{0}{(i)}\gets 0$ for each $i\in[m+1]$ \Comment Initialization ($T_{l}$ is $0$ for $l=0$)
			\For{$l \gets 1$ to $k$} \Comment Solution set is of size at most $l$
			\For{$i \gets 2$ to $m+1$}
			\State $max \gets -\infty$
			\For{$j \gets 1$ to $i$}
			\If{$x(q_j)\le x(q_i)$ and $T_{l-1}(j)+\rho(i,j) > max$}
			\State $max \gets T_{l-1}(j)+\rho(i,j)$
			\EndIf
			\EndFor
			\EndFor 
			\State $T_{l}(i) \gets max$
			\EndFor
		\end{algorithmic}
	\end{algorithm}
	\begin{lemma}\label{lem:algo-simple}
		Algorithm~\ref{alg:simple} solves the maxDominance Problem in $O(km^2 + n\log m)$-time and $O(n+m^2)$-space
	\end{lemma}
	Proof:
		The correctness proof of Algorithm $2$ directly follows from Lemma $2$ of \cite{Choi21}. The Lemma is given below for completeness.
		\begin{lemma}
			(\textit{Lemma~$2$} in \cite{Choi21}) For each $l\in\{1,2,...,k\}, i\in\{1,...,m+1\}$ and $j=\{1,2,..,i\}$, we have
			$$T_{l}(i) = \max\{ S_{l}(i,j): q_j \in Q_{i}^{\lrcorner}\}$$
			$$S_l(i,j) = T_{l-1}(j) + w_i(dom(q_j))$$
		\end{lemma}
		where $Q_{i}^{\lrcorner}$ denotes the set of the points of $Q$ contained in the region $(-\infty, x(q_i)]\times [y(q_i), +\infty)$, and $w_i(dom(q_j)$ denotes the sum of the weights of the points of $P_i$ present in the region $(-\infty, x(q_j)]\times (-\infty, y(q_j)]$. 
		
		Steps $1-5$ takes $O(m+n+m\log m)$ time. We spend $O(n\log m + m^2)$ time for computing $\rho(i,j)$ values for all $j<i$. The time taken by steps $7-18$ is $O(km^2)$. Thus, the total time complexity of Algorithm $2$ is $O(km^2 + n\log m)$. The space used by the algorithm is $O(n+m^2)$.$\square$
	
	\subsection{Further Improvements}
	Instead of explicitly storing $\rho(i,j)$ for all $j<i$, we can implicitly compute them using prefix sums.
	For each $i\in[m]$, we compute prefix sums $psum(i,l)=\sum_{t=1}^{l}w(C_{it})$, for all $l\le i$; the total time used for computing all the prefix sums is $O(m^2)$. 
	We store the prefix sum $psum(i,j)$ if the corresponding cell has non-zero weight, i.e., $w(C_{ij})\ne 0$. Hence, the space used to store the prefix sums is $O(\min\{n, m^2\})$.
	Given a pair of indices $i$ and $j$ such that $j<i$, we can compute $\rho(i,j)$ in $O(1)$-time using the recurrence relation~$(1)$ and the stored prefix sums. Thus, the space-bound reduces to $O(n+m)$.
	\begin{theorem}\label{theorem:algo-simple}
		Given a set of $P$ of $n$ weighted points, a set $Q$ of $m$ unweighted points, and a positive integer $k$, the maxDominance problem can be solved using $O(km^2 + n\log m)$-time and $O(n+m)$-space.
	\end{theorem}
	The algorithm is simple and does not use any complicated data 
	structures.
	However, only for the case when $m=o(\sqrt{n})$, the algorithm is faster than the existing algorithm (Choi et al. \cite{Choi21}).
	\section{Conclusion}
	
	We proposed two algorithms for the \textit{maxDominance} problem. The first algorithm runs (asymptotically) faster as compared to the existing algorithm \cite{Choi21} in the case when $m=o(\sqrt{n})$. For the case when $m=\Omega(\sqrt{n})$, the bounds of our algorithm matches with the existing ones. 
	The other proposed algorithm is simple and easy to explain. Again, its bounds are interesting only in the case when $m=o(\sqrt{n})$.

	\bibliographystyle{abbrv}

\end{document}